# A Security Architecture for Mobile Wireless Sensor Networks

Stefan Schmidt[1], Holger Krahn[2], Stefan Fischer[1], and Dietmar Wätjen[3]

[1] TU Braunschweig, Institute of Operating Systems and Computer Networks,
Mühlenpfordtstr.23, 38106 Braunschweig, Germany
`{schmidt|fischer}@ibr.cs.tu-bs.de`
[2] TU Braunschweig, Institute of Software Systems Engineering,
Mühlenpfordtstr.23, 38106 Braunschweig, Germany
`krahn@sse.cs.tu-bs.de`
[3] TU Braunschweig, Institute of Theoretical Computer Science,
Mühlenpfordtstr.23, 38106 Braunschweig, Germany
`waetjen@iti.cs.tu-bs.de`

**Abstract.** Wireless sensor networks increasingly become viable solutions to many challenging problems and will successively be deployed in many areas in the future. However, deploying new technology without security in mind has often proved to be unreasonably dangerous. We propose a security architecture for self-organizing mobile wireless sensor networks that prevents many attacks these networks are exposed to. Furthermore, it limits the security impact of some attacks that cannot be prevented. We analyse our security architecure and show that it provides the desired security aspects while still being a lightweight solution and thus being applicable for self-organizing mobile wireless sensor networks.

## 1 Introduction

Wireless sensor networks are increasingly showing viable solutions to many challenging problems that require the monitoring of real-world events and are predicted to affect our future daily lives in important ways [8]. Large numbers of small, severely resource-constraint devices form these networks, by cooperating to achieve a common goal, which would be unattainable for the individual nodes.

These networks are usually deployed in uncontrollable environments that are not trustworthy. In addition to common threats in wireless networks, e.g. information disclosure, message injection, and replay attacks, sensor networks are physically accessible and consequently more vulnerable. An attacker may capture and compromise a node and thus be able to control a valid member of the network. Furthermore, a variety of Denial of Service attacks are possible in sensor networks.

In this paper we present a security architecture for self-organizing mobile wireless sensor networks that addresses most of the problems above. It utilizes lightweight cryptographic algorithms that allow for easy authentication between the mobile sensor nodes and secure the communication inside the network. Furthermore, it minimizes the effects of compromised sensor nodes.

The rest of the paper is structured as follows. In the beginning we set the stage for our work by introducing a sensor network setting and its environment. Furthermore, we provide a detailed threat analysis for this setting. Thereafter, we describe the technical aspects of our security architecture and present some implementation details. Subsequently, we provide an analysis of our security architecture. A discussion of related as well as future work and a summary of our contribution conclude the paper.

## 2 Sensor Network Architecture and Environment

In this section we want to introduce the sensor network characteristics on which our security architecture is based. Three groups of aspects have a direct impact on the design of our security architecture: the sensor nodes characteristics, the network characteristics, and the environment.

### 2.1 Sensor Nodes

The sensor nodes are characterised as severely resource-constraint devices in terms of available energy, memory, and computational power. For example, our research node, the Embedded Sensor Board ESB 430/1, developed by the Freie Universität Berlin [19], is powered by three standard AAA batteries, incorporates 60 kbyte flash memory, 2 kbyte RAM, and 8 kbyte EEPROM, and uses the micro controller MSP 430 from Texas Instruments. Furthermore, the sensor nodes are not tamper-proof, due to cost factors and the general difficulty in building such devices [1]. Consequently, it is possible to physically manipulate the devices if captured. For interaction purposes, the nodes are equipped with radio frequency communication capabilities. However, this wireless communication provides only limited bandwidth.

These sensor node-specific factors set several constraints for the security architecture. Since only a fraction of the total memory may be used by the cryptographic algorithms and key material, the security architecture demands very lightweight cryptographic algorithms with relatively short key sizes. Furthermore, cryptographic computations need to be executable in an appropriate amount of time as the execution of cryptographic algorithms is not the main task of the nodes. Due to the limited bandwidth and communication being the most expensive operation in terms of energy, messages should not be extended significantly in length when applying security services.

### 2.2 Sensor Network

The sensor network consists of numerous mobile sensor nodes. It does not incorporate an infrastructure or any kind of hierarchy. In fact, the sensor nodes are equal devices in terms of the role they can play in the network and should self-organize to accomplish their appointed task, without any external guidance or supervision. Thus, the sensor nodes gather information based on their sensing capabilities and make decisions based upon the gathered data. The communication paradigm is based on a distributed virtual shared information space (dvSIS) combined with content-based forwarding of information as described in [6, 14]. The dvSIS describes the state of both the network and the environment. It is based on the information that is acquired by the sensor nodes. Information exchange is based on one-hop communication, which eliminates the need for routing mechanisms. Basically, all information is flooded inside the sensor network and the nodes decide whether or not to forward the received information based on filtering rules, e.g. the information being new or already known from their local dvSIS. However, no device has complete knowledge (i.e. has all information in its dvSIS) and must therefore rely on partial information. Nevertheless, employing these concepts, any node may become a gateway to an external network or observer, which is preferable in mobile networks since the user might not have control over the position of potential gateway-nodes. Another important point in sensor networks is the limited lifetime of sensor data. Sensor data and accordingly events that are derived from it should be communicated in realtime.

The network characteristics, similar to the node characteristics, determine important aspects of the desired security architecture. Considering the node mobility, authentication and key exchanges must not depend on numerous extra messages, since the topology is subject to frequent change. Additionally, all necessary cryptographic functions and key material must reside and be executable on the nodes. With respect to the realtime property of sensor networks, cryptographic algorithms should also be as fast as possible. Finally, the security architecture needs to be scalable to accommodate high numbers of mobile nodes.

### 2.3   Environment

The environment of these sensor networks depends on the assigned task and must in most cases be seen as uncontrollable and not trustworthy. Even scenarios that are characterised by a hostile environment can be envisioned.

This strongly emphasizes that a security architecture must be fault tolerant and ensure certain levels of security even in the case of compromised nodes.

## 3   Threat Analysis

Any sound security architecture requires a thorough threat analysis beforehand to enable an evaluation of its benefits. This section provides a detailed threat analysis for the sensor network that was outlined above.

The commonly known two categories of active and passive attacks can also be applied to sensor networks. However, sensor networks are susceptible to more attacks than ordinary networks, as we will show in the following.

### 3.1   Passive Attacks

Passive attacks on the sensor network are relatively easy because of wireless communication. The illegitimate disclosure of information, which breaks confiden-tiality, presents the major threat here. Therefore, information that is exchanged between the nodes should be encrypted to ensure its confidentiality. Traffic analysis presents a minor threat to this kind of sensor networks since the sensor nodes communicate frequently to publish their sensor readings, derived events, or to coordinate their behaviour.

### 3.2 Active Attacks

Successful active attacks allow the attacker to seriously disrupt the functioning of the network. Several points of attack are imaginable to influence the network in its data capture and decisions. It must be ensured that it is impossible to inject or replay messages into the network. Otherwise an attacker could masquerade as a legitimate member of the network and send its own information or replay old data, which might lead to wrong decision making inside the network. Another means of influencing the network originates from the way sensor networks operate. If an attacker is able to generate physical stimuli he can in parts specify what data the sensor nodes collect. The commonly known threat of message modification, however, does not pose a threat to our sensor network. Due to the local broadcast of the messages, it should be almost impossible to intercept and modify a single message before any other node receives it.

A serious vulnerability arises from the physical accessibility of the sensor nodes. Communicated location information might lead attackers to find out about the positions of single nodes, which further emphasizes the need to keep the communication between the nodes confidential. Furthermore, considering that the sensor nodes are not tamper-proof, it is possible to physically compromise captured nodes. In particular, the attacker is likely to attain a node's cryptographic key material. If an attacker successfully compromises a node, he gains full control about it and might use it to spy on the the rest of the sensor nodes. He is now able to decipher any encrypted communication directed at that particular node and to send legitimate custom messages.

Also several possibilities exist to start a Denial of Service (DoS) attack against a sensor network [22]. For example, an attacker who is able to create physical stimuli might flood the network or at least parts of it, so that real events drown in the artificial noise of stimuli. Furthermore, attackers who possess the location information of the nodes could capture or destroy the nodes one by one. The traditionally known jamming of the wireless medium or sensor specific DoS-attacks such as sleep deprivation torture [21] present threats as well.

## 4 Security Architecture

Generally, asymmetric as well as symmetric cryptography could be employed to achieve security. However, an implementation [12] of an elliptic curve, one of the fastest available asymmetric methods with an underlying field of about $2^{128}$ elements, shows, that it seems to be infeasible to implement a fast public key system on the given nodes. The calculations would delay the message transmis-sion more than a second, which is unacceptable. Consequently we refrain from using asymmetric cryptography.

Our security architecture is based on three different interacting phases: a pairwise key agreement to provide authentication and the initial key exchange, the establishment of sending clusters to extend pairwise communication to broadcast inside the communication range, and encrypted and authenticated communication of sensor data.

### 4.1   Pairwise Key Agreement

To achieve pairwise key agreement we use the Blundo-et-al.-scheme [5], which is based on a predistribution scheme by Blom [4] and enables two nodes to determine a pairwise secret that is solely shared among these two nodes. The scheme is unconditionally secure and resistant against collusion of a maximum of $t$ users. In terms of a cooperative sensor network, this means that an attacker has to compromise more than $t$ nodes to compromise the whole network.

The scheme requires a certificate authority (CA) which randomly generates a symmetric bivariate polynomial $f(x,y)$ of degree $t$ over an arbitrary finite field.

$$f(x,y) = \sum_{i,j=0}^{t} a_{ij} x^i y^j \quad (a_{ij} = a_{ji})$$

The CA, ensuring that each sensor has an unique *ID*, evaluates the polynomial in the following way:

$$g_{ID}(x) = f(x, ID).$$

Thus $g_{ID}$ is a polynomial of degree $t$ with a single variable $x$. The CA transfers the individual key material (the coefficients of the polynomial $g_{ID}$) to the nodes prior to the deployment of the sensor nodes.

Two nodes are able to determine their pairwise secret by evaluating their private polynomial $g_{ID}(x)$ where $x$ denotes the other node's identity. It can be derived directly from the symmetry of the polynomial $f(x,y)$ that both nodes calculate the same value.

An efficient way to implement the Blundo-et-al.-scheme is presented in appendix A.

### 4.2   Sending Clusters

To communicate securely, every node establishes a randomly generated key within its neighbourhood. This key is used solely by this node to encrypt and authenticate its messages. If a node receives a message of which the content cannot be decrypted and authenticated (i.e. it does not know the key) it calculates the pairwise secret for the sender and itself using the Blundo-et-al.-scheme. This secret is then used to transfer its own key secretly to the sender, which replies by transferring its key. The node automatically sends its own key, because we assume that if it could not decrypt and authenticate the other message, a change in the topology must have happened and the other node equally does not know its sending key.

This protocol enables a node to establish its key in a new environment and get to know the other nodes' keys with just two messages (request and reply) per neighbour. It is not restricted to an initialisation phase and can be used at any time and with any other legitimate node of the network. Thus it supports mobile sensor networks as well as the deployment of new nodes at a later point of time.

### 4.3 Encrypted and Authenticated Communication

The sensor nodes always broadcast messages to their direct neighbourhood. For encryption purposes the counter mode of operation [10] (see also figure 1) is used, which allows an encryption of the message without changing its length. In addition to the message itself the counter $s_j$ is added, which results in ordered and unique messages. This overhead can be avoided if both sender and receiver increment the counter after each transmission. However, due to the lossy nature of the communication in sensor networks this procedure seems inappropriate. A message loss by any of the neighbours would require two additional messages (request of counter value and reply). Thus, the counter value is included in every message. It is not necessary to use the full block size of an encryption algorithm

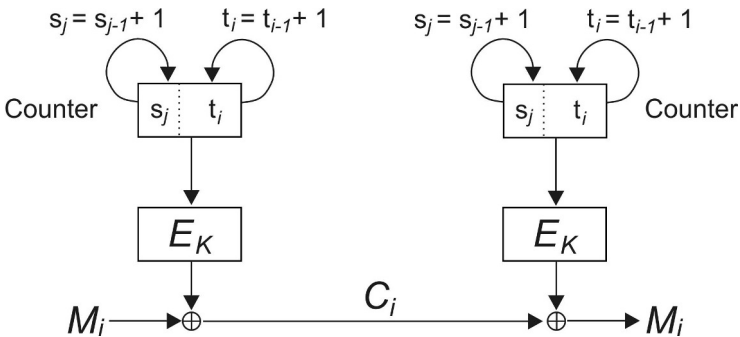

**Fig. 1. Counter mode of operation.** The counter register is divided into two parts: $s$ which is incremented once per message and $t$ which is reset to zero for a new message and incremented once per block

for the counter value, because any counter length can be padded with zeros. Since the same counter should not be used twice, the size of the counter limits the number of messages which can be encrypted by a single key.

Our security architecture can be implemented using any encryption algorithm. Often the RC5 algorithm [18] is suggested as being a well suitable algorithm for sensor networks. We give two reasons why this choice is not optimal for our architecture. First, the key expansion step of the algorithm cannot be integrated in the encryption process. This results in the need to have round keys stored for the whole encryption process which leads to a slightly higher memory requirement than desired (at least 112 bytes for RC5-32/12/8 when the original key needs to be kept). Second, the RC5 algorithm extensively uses circular shifts. This operation is often not supported by low cost processors and must be emulated by single step shifts, which leads to bad runtime behaviour.

For our existing implementation we examined the AES-final candidates. The Rijndael algorithm [16] seems to offer the best performance, but requires large lookup-tables. Also this algorithm is not constructed for a key length shorter than 128 bit, which are often used in sensor networks. Simply padding the key might expose new flaws. Therefore, we decided to incorporate Serpent [2] which has good runtime behaviour because it can be implemented using logical operations only. These are much faster than circular shifts on low cost platforms. A reduction of the number of rounds from 32 to 16 yields to linear speed-up while at the same time not allowing any published attacks to be successful.

The encryption algorithm can be reused for the well-know CBC-MAC. The result of the chain of encryption operations can be used to ensure the integrity and the authentication of sensor data. Once again, there is no need to add the complete output of the MAC-function as a checksum to the message, because 16 byte of overhead per message seems to be inappropriate for sensor networks.

## 5   Implementation

We implemented a prototype of our security architecture using the sensor nodes ESB 430/1 [19]. The algorithms require 17,1 Kb. This results mainly from the Serpent algorithm being a standard implementation, which has only been optimized with regard to volatile memory usage and speed but not code size. During run-time the algorithms need additional 86 bytes for their operations. The average delay caused by this is 30,9 ms per 16 byte of cleartext information.

The security degree $t$ (number of nodes an attacker has to compromise to successfully calculate the CA's polynomial) is completely adjustable but influences the run-time and memory requirement. For our 80-bit implementation (size of the pairwise secrets) $(t+1) \cdot 10$ bytes are required. The coefficients can be freely distributed between volatile memory, code space and EEPROM wherever space is available. For our testing, we chose the security degree to be 20 and stored the coefficients in memory. Additionally, a node has to store the sending key of each neighbour. In our implementation nodes store up to 20 sending keys. After that they start to reuse the space by deleting the oldest one. In summary, we use around a quarter of the available memory for our security architecture.

Also, for our prototype we used a 4 byte MAC and a 2 byte index which merely results in an extension of the messages by 6 bytes.

## 6   Analysis

By employing our security architecture we are able to prevent most of the depicted threats while considering the constraints that the sensor network de-mands. It provides confidentiality as well as integrity for the communicated information and ensures the authenticity of the sensor nodes. Furthermore, it minimizes the effects of compromised nodes. All deployed cryptographic algorithms are efficient in terms of run-time and memory usage and do not extend the messages significantly.

Confidentiality is achieved by encrypting the messages. This prevents any illegitimate disclosure of information. Futhermore, the CBC-MAC ensures the integrity of the messages. We use short MACs in our architecture to reduce the message overhead, which, on the one hand, enables the attacker to determine legitimate messages by brute force. On the other hand, he has no control over the message content due to the encryption of the message. Furthermore, the counter mode of operation, which is used in the encryption process, automatically adds an index to each message without adding overhead compared to simple encryption. Hereby, replay attacks are prevented, because a receiver can keep track of already used counter values, while the messages are only slightly extended.

Only legitimate nodes can join the communication, since no illegitimate node is able to derive the sending key of the other nodes. An attacker may conduct the pairwise key agreement protocol using a captured key request message from another node. However, it is only possible for him to attack the encrypted sending key (encrypted with the pairwise secret key), which, if no new ways to attack Serpent become known, can be seen as impossible. Furthermore, the authentication process is implicitly included in the key exchange, which only requires two additional messages.

One threat we cannot prevent is the possible capturing and compromising of nodes. However, our security architecture minimizes its effects. While an attacker, who successfully compromised a node, may be able to authenticate himself to the network and be able to join in the communication, he can only do so using the one compromised node. The communication between non-compromised nodes remains secure. This holds true for up to $t$ compromised nodes, which is based on the security specified by the initial symmetric bivariate polynomial calculations. It is important to note that $t$ is not neccessarily linearly dependent on the actual network size. It denotes the number of nodes that need to be physically compromised, which is by far the most expensive attack in this scenario and thus more unlikely than radio-based attacks.

Due to the fact that cryptographic algorithms cannot prevent DoS attacks, the security architecture does not address these threats any further.

Our architecture is scalable and robust since all operations take place inside a node's communication radius. Thus, the actual size of the network does not influence its local security associations.

## 7   Related Work

Security in sensor networks has been studied by several other researchers.

Perrig et al. developed the security architecture SPINS, which is based on the two protocols SNEP, a protocol for data confidentiality, two-party data authen-tication, and data freshness and µTESLA, a broadcast authentication protocol [17]. Their architecture relies on the concept, that every node shares a secret key with a trusted base station, which is at all times able to communicate with every node in the network.

Furthermore, several key management schemes have been put forward for sensor networks: Basagni et al. proposed a solution to periodically update a symmetric key which is shared by all nodes in the network [3]. Their solution is based on the assumption that all nodes are constructed tamper-proof, which is not always the case [1]. Carman et al. studied several key management protocols in sensor networks with respect to performance on different hardware platforms [7]. Zhu et al. proposed the Localized Encryption and Authentication Protocol (LEAP) [24] which utilizes four types of keys for each node. These are used for different purposes and range from the individual key that is shared with the base station, up to a group key that is shared with all nodes in the network. Eschenauer and Gligor presented a pool-based random key predistribution system [11], which Chan et al. extended by presenting three new mechanisms for key establishment [9].

The key setup algorithm that is most related to ours is presented by Liu and Ning [15]. They proposed a combination of the Blundo-et-al.-scheme and the pool-based random key predistribution system by Eschenauer and Gligor [11]. With the same storage requirements, the CA uses shorter polynomials than the non-modified Blundo-et-al.-scheme but deploys more than one private polynomial from a pool to a single node. If two nodes share at least one polynomial from the same scheme, they use this to determine the pairwise secret as described in section 4.1. The combination has a higher resistance against the possibility of compromising nodes than the Blundo-et-al.-scheme if the compromised nodes are picked randomly, with the disadvantage that only certain nodes are able to communicate. Lui and Ning state that this can be amended by using an intermediate node to establish a connection between two nodes which have no scheme in common. We think this enables new attacks in mobile sensor networks. A single compromised node is able to vouch for other nodes the attacker has fabricated and allows them to become part of the sensor network.

Furthermore, the system is only advantageous if an attacker is not able to determine the identity of the polynomials stored on a certain node. Therefore, in [11] a protocol is proposed where an encrypted list, using the pairwise keys for the encryption, is exchanged. Adding to the disadvantage that this requires an additional message exchange of at least 32 byte length (64-bit-block cipher and 4 polynomials per node), the system is still not secure against compromising nodes. If an attacker has compromised at least a single node, he can actively use it to communicate with others to determine whether they share any polynomial. With this method, the attacker is able to actively attack nodes that possess certain polynomials. Due to the longer messages and the described attacks, we decided to use the unmodified Blundo-et-al.-scheme.

Wood and Stankovic identified several DoS attacks in sensor networks [22] and presented a protocol, which allows to map regions that are subject to DoS by radio jamming [23].

## 8  Summary and Future Work

In this paper, we have proposed a security architecture that provides confidentiality, integrity, and authentication for a mobile wireless sensor network. For this purpose, we have presented algorithms to easily set up pairwise secret keys between the mobile sensor nodes and to establish a sending cluster per node, in which it can communicate its messages securely. Furthermore, our solution minimizes the effects of compromised nodes. Compromising an adjustable number of sensor nodes does not compromise the whole security architecture but restricts the security breach to the immediate neigbourhood of the compromised node. Finally, we have implemented a prototype of our security architecture, which clearly shows that it is a lightweight solution and applicable for self-organizing mobile wireless sensor networks.

Several directions for future research arise from our solution. First, we intend to simulate our approach, using NS-2, in order to determine the maximum grade of node-mobility our security architecture is able to cope with. Second, we would like to integrate the ability to identify compromised nodes and methods to exclude them from the network. Another interesting question is to determine how much further we can optimize the employed algorithms with respect to memory usage and speed.

## A    Implementation Details

This part illustrates a way to efficiently implement the Blundo-et-al.-scheme. Two concepts are combined: first, the Horner Rule minimizes the multiplication effort and second, the choice of the field determines the speed of the modular reduction.

For an efficient implementation the *ID*'s can be reduced to a certain range, e.g. $0 < ID < 2^{16}$. If the evaluation is done using the Horner Rule (e.g. [13], see also figure 2) only multiplications between a short *ID* and a longer field element have to be done, which allows for a much faster modular arithmetic than the general case.

$$a_n ID^n + a_{n-1} ID^{n-1} + \ldots + a_0 =$$
$$(\ldots (a_n ID + a_{n-1}) ID + a_{n-2}) ID + \ldots) + a_0$$

**Fig. 2.** Horner Rule

The choice of the field $F$ determines the range and the speed of the evaluation of the common secret. The scheme can be implemented using any field. Consequently, a field, which suits the existing hardware platform well, should be chosen. The arithmetic of $GF(p)$ is supported by the existing hardware multiplier of the MSP430 processor. Therefore, this choice is faster than an implementation using $GF(2^m)$.

The following example assumes that the identities are limited to $0 < ID < 2^{16}$ and illustrates the usage of a 80-bit Generalized Mersenne Prime, which allows for an efficient modular arithmetic [20], $p = 2^{80} - 2^{64} - 2^{32} - 1$. First a normal multiplication of a 16-bit identity and a 80-bit coefficient is calculated. Afterwards one can rearrange the 96-bit result $r = \sum_{i=0}^{5} r_i 2^{16i}$ to $s = \sum_{i=0}^{4} r_i 2^{16i}$ and $t = 2^{64} r_5 + 2^{32} r_5 + r_5$. From the special form of $p$ follows that $s + t = r \mod p$. The advantage is that the reduction modulo $p$ can be calculated by a single 80-bit addition and at most 2 80-bit subtractions. This is substantially faster than any standard method that assumes two equally long operands.